\documentclass[12pt,a4paper]{article}
\pdfoutput=1
\usepackage[utf8]{inputenc}
\usepackage{amsmath}
\usepackage{amsfonts}
\usepackage{amssymb}
\usepackage{graphicx}
\author{Xianglin Liu \\
{\small \textit{Department of physics, Carnegie Mellon University} }\\
Yang Wang \\
{\small \textit{Pittsburgh supercomputing center, Carnegie Mellon University}}\\
Markus Eisenbach  \\
{\small \textit{Center for Computational Sciences, Oak Ridge National Laboratory} }\\
G. Malcolm Stocks \\
{\small \textit{Materials science and technology division, Oak Ridge National Laboratory} }
}

\newcommand{\mr}{\mathbf{r}}

\begin{document}
\title{A full-potential approach to the relativistic single-site Green's function \footnote{\footnotesize{This manuscript has been authored by UT-Battelle, LLC under Contract No. DE-AC05-00OR22725 with the U.S. Department of Energy.  The United States Government retains and the publisher, by accepting the article for publication, acknowledges that the United States Government retains a non-exclusive, paid-up, irrevocable, world-wide license to publish or reproduce the published form of this manuscript, or allow others to do so, for United States Government purposes.  The Department of Energy will provide public access to these results of federally sponsored research in accordance with the DOE Public Access Plan(http://energy.gov/downloads/doe-public-access-plan).}}}
\maketitle
\begin{abstract}

One major purpose of studying the single-site scattering problem is to obtain the scattering matrices and differential equation solutions indispensable to  multiple scattering theory (MST) calculations. On the other hand, the single-site scattering itself is also appealing because it reveals the physical environment experienced by electrons around the scattering center. In this paper we demonstrate a new formalism to calculate the relativistic full-potential single-site Green's function. We implement this method to calculate the single-site density of states and electron charge densities. The code is rigorously tested and with the help of Krein's theorem, the relativistic effects and full potential effects in group V elements and noble metals are thoroughly investigated.

\end{abstract}

\section{Introduction}
The multiple-scattering theory (MST) is a DFT based ab-initio method that is widely applied to the calculation of the electronic structure of metals, alloys and impurities. One crucial step in MST calculation is to solve the single-site scattering problem. Combined with the position information of the atoms, solutions of the single-site scattering problem can be used to construct the Green's function of the whole system, from which most physical quantities can be extracted. Because of the essential role played by the single-site scattering in MST, there have been a constant effort to improve it in the last few decades \cite{Ebert}. The earliest MST solves the Schr\"odinger's equation and employs the muffin-tin potential approximation, i.e. the potential is spherically symmetric within a bounding sphere and is constant outside.  Then the generalization to full potential (FP)\cite{Zeller} and relativistic cases \cite{Feder,StrangeA,StrangeB,Jenkins}, and eventually  combination of the two, i.e. relativistic full potential (RFP),  were proposed and implemented using various schemes \cite{Tamura,XDWang,Lovatt,Freeman}. Among them one widely used RFP MST code is developed by Huhne et al \cite{Huhne} to either solve the coupled integral equations iteratively using Born's series expansion or to directly solve the coupled differential equations. In a recent paper\cite{Geilhufe}, the coupled differential equations are also solved by matching the regular solutions at the boundary of the Wigner–Seitz cell.

In this paper we present a alternative formalism to tackle the RFP single-site scattering problem by directly solving the Dirac differential equation.
This method is a relativistic generalization of the non-relativistic FP MST method\cite{Rusanu}. Compared to other RFP methods, this new formalism has the feature that the differential equation, the $t$ matrix and the single-site Green's function are all expressed in terms of the $r$ dependent sine and cosine matrices. We would like to point out that the sine and cosine matrices technique to obtain the solutions for relativistic scattering theory was first proposed by X. D. Wang \textit{et al.} \cite{XDWang}, based on the phase integral technique\cite{Williams},  and later implemented by S. B. Kellen and A. J. Freeman \cite{Freeman}. The major difference between our method and the Kellen \& Freeman’s approach is in the calculation of observables. They find the observables by searching for energy eigenvalues and eigenfunctions of the KKR secular equation, while in our case all physical quantities are expressed in terms of the Green's functions, without the compute-intensive eigenvalue searching and the wavefunction orthonormalization procedures.

As a test of our code, the single-site DOS are calculated using both the Green's function method and the Krein's theorem method and two results are compared. To investigate the relativistic effects and the full potential effects,  we study the density of states of noble metals and group V elements. Finally, the charge density of tantalum is calculated and some interesting relativistic features are discussed.

\section{Theory}
In the following discussion, we use the atomic Rydberg units; i.e., $\hbar=1, m=1/2$, $c\approx 274$, etc. To construct the Green's function, we first need to solve the single-particle Dirac equation:
\begin{align}
\{-c\boldsymbol{\alpha} i \nabla+\beta mc^2+V(\mr)\}\psi(E, \mr)=W  \psi(E,\mr), \label{Dirac}
\end{align}
where $\boldsymbol{\alpha}$, $\beta$ are the Dirac matrices, $W$ and $E$ are the relativistic energy and relativistic kinetic energy, respectively. $W$ and $E$ are related to momentum $p$ by
\begin{align}
 W&=\sqrt{(mc^2)^2+(pc)^2},\\
 E&=W-mc^2.
 \end{align}
The solution $\psi(E,\mr)$ is a Dirac spinor. Within the framework of spin polarized relativistic density functional theory \cite{SPDFT}, the effective potential is written as
\begin{align}
V(\mr) =
\begin{pmatrix}
v(\mr) + \boldsymbol{\sigma} \cdot \mathbf{B}(\mr)  & 0  \\
0 & v(\mr) -\boldsymbol{\sigma} \cdot \mathbf{B}(\mr) \\
\end{pmatrix} ,
\end{align}
where $\boldsymbol{\sigma} $ are the Pauli matrices and $\mathbf{B}$ are effective magnetic fields. In the following discussion we will focus on nonmagnetic systems and discard the $\mathbf{B}$ terms. For $r>R_c$, where $R_c$ is the radius of the circumscribed sphere of the Wigner-Seitz cell, the effective potential $V(\mr)$ vanishes, and the solutions of the corresponding free-space Dirac equation are well known. The right-hand solutions are
\begin{align}
J_{\Lambda}(E, \mr)&=\left( \frac{W+mc^2}{c^2}\right)^{1/2} \left(
\begin{array}{c}
j_l(pr)\chi_\Lambda(\hat{\mathbf{r}})\\
\frac{ipc S_\kappa}{W+mc^2} j_{\bar{l}}(pr) \chi_{\bar{\Lambda}}(\hat{\mr})\\
\end{array}
\right), \label{J}\\
N_{\Lambda}(E, \mr)&=\left( \frac{W+mc^2}{c^2}\right)^{1/2} \left(
\begin{array}{c}
n_l(pr)\chi_\Lambda(\hat{\mathbf{r}})\\
\frac{ipc S_\kappa}{W+mc^2} n_{\bar{l}}(pr) \chi_{\bar{\Lambda}}(\hat{\mr})\\
\end{array}
\right), \label{N}
\end{align}
where $j_l(pr)$ and $n_l(pr)$ are the usual spherical Bessel functions of the first kind and the second kind, with angular momentum index $l$. $\Lambda$ stands for the pair of relativistic angular-momentum indices $(\kappa,\mu)$ and $S_{\kappa}$ is the sign of $\kappa$ index.  $\bar{\Lambda}=(-\kappa,\mu)$  and
\begin{align}
\bar{l} =
\begin{cases}
 l+1    & \quad \text{if } \kappa < 0\\
 l-1    & \quad \text{if } \kappa > 0\\
\end{cases}.
\end{align}
$\chi_{\Lambda}$ and $\chi_{\bar{\Lambda}}$  are the spin-angular functions
\begin{align}
\chi_{\Lambda}(\hat{\mr})=\sum_{m_s=\pm1} C(l,j,\frac{1}{2}| \mu-m_s,m_s) Y_{l,\mu-m_s}(\hat{\mr}) \phi_{m_s}, \\
\chi_{\bar{\Lambda}}(\hat{\mr})=\sum_{m_s=\pm1} C(\bar{l},j,\frac{1}{2}| \mu-m_s,m_s) Y_{\bar{l},\mu-m_s}(\hat{\mr}) \phi_{m_s},
\end{align}
where $C(l,j,\frac{1}{2}| \mu-m_s,m_s)$ are the Clebsch-Gordan coefficients, $Y_{l,\mu-m_s}$ are spherical harmonics and  $\phi_{m_s}$ are Pauli spinors
\begin{align}
\phi_{1/2}=\left(
\begin{array}{c}
1\\0
\end{array}
\right),
\;\;\;  \;\;\;
\phi_{-1/2}=\left(
\begin{array}{c}
0\\1
\end{array}
\right).
\end{align}
In addition to the right-side solutions,  we also need the left-hand solutions to construct the Green's function (see Appendix A and B).  In free-space, the left-hand solutions are given by
\begin{align}
J^+_\Lambda(E, \mr)=&\left( \frac{W+mc^2}{c^2}\right)^{1/2} \left( j_l(pr)\chi_\Lambda^\dagger (\hat{\mathbf{r}}), \;\; \frac{-ipc S_\kappa}{W+mc^2} j_{\bar{l}}(pr) \chi_{\bar{\Lambda}}^\dagger (\hat{\mr}) \right), \label{Jplus}\\
N^+_\Lambda(E, \mr)=&\left( \frac{W+mc^2}{c^2}\right)^{1/2} \left( n_l(pr)\chi_\Lambda^\dagger (\hat{\mathbf{r}}), \;\; \frac{-ipc S_\kappa}{W+mc^2} n_{\bar{l}}(pr) \chi_{\bar{\Lambda}}^\dagger (\hat{\mr}) \right). \label{Nplus}
\end{align}

For $r<R_c$, the Dirac equation solutions can only be obtained numerically. By using the phase integral technique\cite{Williams} , the solutions are written in terms of the free-space solutions as (see Appendix A)
\begin{align} \label{eq03}
\psi_\Lambda(E, \mr)=\sum_{\Lambda'} \{ S_{\Lambda' \Lambda}(E, r) N_{\Lambda'}(E, \mr) -C_{\Lambda' \Lambda }(E, r) J_{\Lambda'}(E, \mr) \},
\end{align}
where the $r$ dependent cosine matrix $ C_{\Lambda' \Lambda }(E, r)$ and sine matrix  $S_{\Lambda' \Lambda }(E, r)$ are defined as
\begin{align}
C_{\Lambda' \Lambda}(E, r)&=p\int_{0<r'<r} d^3\mr' N_{\Lambda'}^+(E, \mr') V(\mr') \psi_\Lambda (E, \mr') - \delta_{\Lambda \Lambda'}, \label{eq04} \\
S_{\Lambda' \Lambda}(E, r)&=p\int_{0<r'<r} d^3\mr' J_{\Lambda'}^+ (E, \mr') V(\mr')\psi_\Lambda (E, \mr'). \label{eq05}
\end{align}
Note that this expression is also valid for $r>R_c$, because $S_{\Lambda' \Lambda }(E, r)$ and $C_{\Lambda' \Lambda }(E, r)$ will become constants outside the circumscribed sphere. To distinguish these constants from $S_{\Lambda' \Lambda }(E, r)$ and $C_{\Lambda' \Lambda }(E, r)$ we denote them as $S_{ \Lambda' \Lambda } (E)$ and $C_{\Lambda' \Lambda }(E)$.
Equations (\ref{eq03}), (\ref{eq04}) and (\ref{eq05}) form a set of coupled integral equations. $C_{\Lambda \Lambda'}( E, r)$ and $S_{\Lambda \Lambda'}(E, r)$ can be obtained by solving the corresponding differential equations
\begin{align}
\frac{d}{dr} S_{\Lambda' \Lambda}(E, r)&=r^2\int d\hat{\mr} \; p J_{\Lambda'}^+(E, \mr)V(\mr) \psi_\Lambda(E, \mr), \label{eq06} \\
\frac{d}{dr} C_{\Lambda' \Lambda}(E, r)&=r^2\int d\hat{\mr}\; p N_{\Lambda'}^+(E, \mr)V(\mr) \psi_\Lambda(E, \mr). \label{eq07}
\end{align}
Note that the integral is only upon the angular part. For regular solutions, the boundary conditions are
\begin{align}
\psi_\Lambda (E, \mr)_{  \mr \to 0} = J_\Lambda (E, \mr).
\end{align}
or equivalently
\begin{align}
S_{\Lambda' \Lambda}(E, 0)&=0, \\
C_{\Lambda' \Lambda}(E, 0)&=-\delta_{\Lambda \Lambda'}.
\end{align}
To see the boundary conditions are well defined, we note the effective electric potential has a spherically symmetric $1/r $ behavior at the origin. As a result the integrals on the right side of equation (\ref{eq04}) and (\ref{eq05}) vanish at the origin because spherical bessel functions with different $l$ indices can not be coupled by spherical potential. 

Next, we proceed to discuss technical details on solving the coupled differential equations. The explicit expressions of the differential equations (\ref{eq06}) and (\ref{eq07}) are given by
\begin{align}
\frac{d}{dr} S_{\Lambda' \Lambda}(E, r)=\sum_{\Lambda''} a_{\Lambda'' \Lambda'}(E, r) S_{\Lambda'' \Lambda}(E, r)-\sum_{\Lambda''} b_{\Lambda'' \Lambda'}(E, r) C_{\Lambda'' \Lambda}(E, r), \label{dSdr}
\end{align}
where
\begin{align}
a_{\Lambda''\Lambda'}(E, r)&=r^2\int d\hat{\mr} \;p J_{\Lambda'}^+(E, \mr)V(\mr) N_{\Lambda''}(E, \mr),  \label{aa}\\
b_{\Lambda''\Lambda'}(E, r)&=r^2\int d\hat{\mr} \;p J_{\Lambda'}^+(E, \mr)V(\mr) J_{\Lambda''}(E, \mr), \label{bb}
\end{align}
and
\begin{align}
\frac{d}{dr} C_{\Lambda' \Lambda}(E, r)=\sum_{\Lambda''} c_{\Lambda'' \Lambda'}(E, r) S_{\Lambda'' \Lambda}(E, r)-\sum_{\Lambda''} d_{\Lambda'' \Lambda'}(E, r) C_{\Lambda'' \Lambda}(E, r), \label{dCdr}
\end{align}
where
\begin{align}
c_{\Lambda''\Lambda'}(E, r)&=r^2\int d\hat{\mr} \;p N_{\Lambda'}^+(E, \mr)V(\mr) N_{\Lambda''}(E, \mr), \label{cc}\\
d_{\Lambda''\Lambda'}(E, r)&=r^2\int d\hat{\mr} \;p N_{\Lambda'}^+(E, \mr)V(\mr) J_{\Lambda''}(E, \mr). \label{dd}
\end{align}
For full potential calculation, the effective potential $V(\mr) $ is expanded in terms of the spherical harmonics
\begin{align}
V(\mr)&=\sum_{L_v} V_{L_v}(r)Y_{L_v}(\hat{\mr}).
\end{align}
The angular integral in equation (\ref{aa}),(\ref{bb}),(\ref{cc}) and (\ref{dd}) can be done analytically and written in terms of the Gaunt coefficients
\begin{align}
C_{L,L'}^{L''}=\int d\hat{\mr} \; Y_L(\hat{\mr}) Y_{L'}^* (\hat{\mr}) Y_{L''} (\hat{\mr}).
\end{align}
In practice, the differential equations are solved on exponential radial grid ${r=r_0\exp(x)}$, using the fourth-order Bashforth-Adams-Moulton predictor-corrector method.

After we obtain the solutions of the Dirac equation, we can use them
to construct the single-site Green's function, which has the following expression (see Appendix B)
\begin{align}
G(E, \mr,\mr')=\sum_{\Lambda \Lambda'}Z_\Lambda(E, \mr) t_{\Lambda\Lambda'}(E) Z^+_{\Lambda'}(E, \mr')-\sum_{\Lambda} Z_\Lambda(E, \mr) \mathcal{J}^+_{\Lambda}(E, \mr')   \label{Green} .
\end{align}
where $\mathcal{J}^+_{\Lambda'}(E, \mr)$ are the solutions of equation (\ref{Dirac}) with the boundary condition that $\mathcal{J}^+_{\Lambda'}(E, \mr) =J^+_{\Lambda'}(E, \mr) $  outside the Wigner-Seitz cell.
$Z_\Lambda(E, \mr)$ is another set of regular solutions defined as
\begin{align}
Z_\Lambda(E, \mr)=p\sum_{\Lambda'} \psi_{\Lambda'}(E, \mr) S^{-1}_{\Lambda' \Lambda}(E). \label{Z}
\end{align}
The $t$ matrix is given by equation (\ref{TInv}) in Appendix B and the explicit expression is
\begin{align}
t_{\Lambda \Lambda'}(E)=-\frac{1}{p} \sum_{\Lambda''} S_{\Lambda \Lambda''}(E) \left( C_{\Lambda'' \Lambda'}\left(E\right)-i\;S_{\Lambda'' \Lambda'}\left( E\right) \right)^{-1}.
\end{align}

From the Green’s function, it's straightforward to calculate physical quantities of the system. Here we focus on  the charge density and the density of states. The first one is necessary to calculate the new potential in a SCF cycle and the latter one is needed to determine the Fermi energy of the system.
The charge density is given by
\begin{align}
\rho(\mr)=-\frac{1}{\pi} \mathrm{Im} \;\mathrm{Tr} \int^{E_F}_0  G(E,\mr,\mr) dE, \label{CD}
\end{align}
where the Fermi energy $E_F$ is chosen to give the correct number of total electrons. The density of states is given by
\begin{align}
n(E)=-\frac{1}{\pi} \mathrm{Im} \; \mathrm{Tr}\int_{\Omega} G(E,\mr,\mr) d\mr, \label{DOS}
\end{align}
where $\Omega$ denotes the wigner seitz cell. For real energy, the imaginary part of the $Z\mathcal{J^+}$ term in equation (\ref{Green}) vanishes, so we can instead write the density of states $n(E)$ and the charge density $\rho(\mr)$ as
\begin{align}
n(E)&=-\frac{1}{\pi} \mathrm{Im} \; \mathrm{Tr}\int_{\Omega} \sum_{\Lambda \Lambda'}Z_\Lambda(E, \mr) t_{\Lambda\Lambda'}(E) Z^+_{\Lambda'}(E, \mr) d\mr, \\
\rho(\mr)&=-\frac{1}{\pi} \mathrm{Im} \; \mathrm{Tr}\int^{E_F}_0  \sum_{\Lambda \Lambda'}Z_\Lambda(E, \mr) t_{\Lambda\Lambda'}(E) Z^+_{\Lambda'}(E, \mr) dE.
\end{align}
The benefit of this expression is that it does not contain the irregular solutions $\mathcal{J}_\Lambda (\mr)$, which are difficult to evaluate precisely near the origin.
Even though the focus of this paper is on single-site Green's function , at the end of this section we would like to make a few comments about the MST Green's function, which has the expression \cite{Faulkner}
\begin{align}
G(E,\mr,\mr')=\sum_{\Lambda \Lambda'}Z_\Lambda(E, \mr) \tau_{\Lambda\Lambda'}(E) Z^+_{\Lambda'}(E, \mr')-\sum_{\Lambda} Z_\Lambda(E, \mr) \mathcal{J}^+_{\Lambda}(E, \mr') .
\end{align}
Compare it with equation (\ref{Green}) we see the only difference is the replacement of the $t$ matrix by the scattering-path matrix $\tau_{\Lambda\Lambda'}$\cite{TauM}, which describes the scattering of the electron along all paths in the solid. The scattering-path matrix can be constructed from the $t$ matrix and the  structure constants  $G_{\Lambda\Lambda'}$, which describe the propagation of free electrons between two atoms. Therefore, it's straightforward to implement our formalism to the MST calculations.
\section{Krein's Theorem}
\begin{figure}[h!]
\centering
    \includegraphics[width=0.45\textwidth]{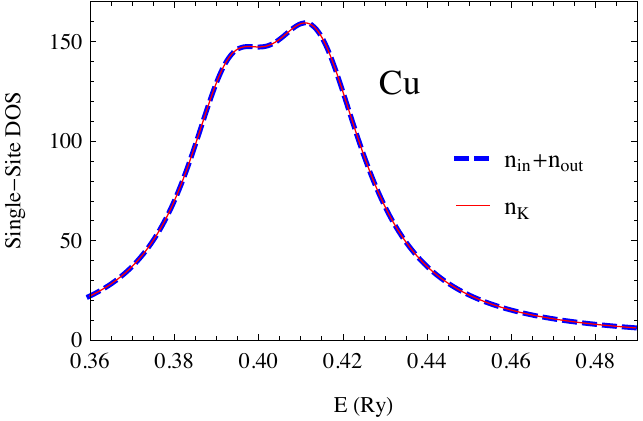}
    \includegraphics[width=0.45\textwidth]{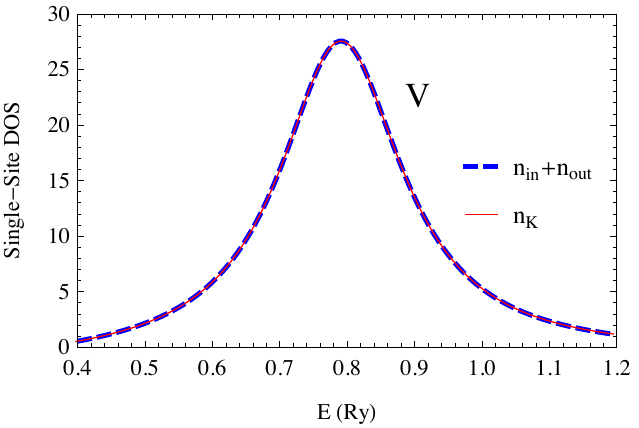}
\caption{ (Color online) Comparison of the DOS from Green's function method and the Krein's theorem method. The blue solid lines show the Krein DOS $n_{\mathrm{K}}$. The dashed lines show $n_{\mathrm{in}}+n_{\mathrm{out}}$ calculated using the Green's function. Because of the good agreement the two lines actually overlap.  } \label{fig:GreenVSKrein}
\end{figure}

For single-site scattering, the density of states can be calculated from either the Green’s function or the Krein’s theorem \cite{Krein}\cite{Sam}. A comparison of the two results will be a good test of our code. First we focus on the Krein’s theorem method. Details of the relation between the Krein’s theorem and the Green’s function in the non-relativistic case has been derived in a previous paper \cite{YangWang} and most of them can also be applied to the relativistic case. Therefore, in this section we will just establish notation and quote known results.
By applying the Krein's theorem to the scattering theory, it has been proved that the integrated density of states (IDOS) is given by
\begin{align}
N_K(E)=-2\xi(E)=N(E)-N_0(E)+n_c, \label{NK}
\end{align}
where $N(E)$ is the single-site density of states, $N_0(E)$ is the free electron integrated density of states, $n_c$  is the total number of core electrons, which is irrelevant here since we are only interested in  valence electrons. $\xi (E)$ is the Krein's spectral shift function \cite{Krein} related to the standard unitary S-matrix $\mathbf{S}(E)$ by \cite{Birman}
\begin{align}
e^{-i 2\pi\xi(E)}=\mathrm{det}  \mathbf{S}(E).
\end{align}
The S-matrix is obtained from the $t$ matrix using the relation 
\begin{align}
\mathbf{S}_{\Lambda\Lambda'}(E)=\delta_{\Lambda \Lambda'}(E)-2 i p t_{\Lambda\Lambda'}(E).
\end{align}
The Krein DOS $n_K(E)$ is obtained by taking the derivative of $N_\mathrm{K}$,
\begin{align}
n_{\mathrm{K}}(E)=\frac{d N_{\mathrm{K}} (E)}{dE}. \label{dNdE}
\end{align}
Note that $n_K(E)$ is the DOS for the entire space. To compare the DOS obtained from the Green's function with $n_K(E)$, we express the DOS inside and outside the region bounded by a sphere of radius $R_c$ in terms of Green's function as
\begin{align}
n_{\mathrm{in}}(E)&=-\frac{1}{\pi}\mathrm{Im}\int_{0<r<{\mathrm{R_c}}} \mathrm{Tr}(G(E,\mr,\mr)-G_0(E,\mr,\mr))  d\mr, \\
n_{\mathrm{out}}(E)&=-\frac{1}{\pi}\mathrm{Im}\int_{r>{\mathrm{R_c}}} \mathrm{Tr}(G(E,\mr,\mr)-G_0(E,\mr,\mr))  d\mr,
\end{align}
and we should have $n_{\mathrm{K}}(E)=n_{\mathrm{in}}(E)+n_{\mathrm{out}}(E)$.  The $G(E,\mr,\mr)$ term in $n_{\mathrm{in}}$ is evaluated numerically and the  $G_0(E,\mr,\mr)$  term  is evaluated analytically using the following expression of the free space Green's function
\begin{align}
\mathrm{Tr}\; G_0(E, \mr,\mr)=&\lim_{\mr'\to\mr}\;\mathrm{Tr}-\frac{1}{c^2}(\mathbf{\alpha}\cdot \mathbf{p}+\mathbf{\beta} mc^2 + W)\frac{e^{ip(\mr-\mr')}}{4\pi (\mr-\mr')} \\
=&-\frac{i p W}{\pi c^2}.
\end{align}
Therefore, the contribution to the DOS from the inside integral is
\begin{align}
n_{\mathrm{in}}(E)=-\frac{1}{\pi}\mathrm{Im}  \int_0^{\mathrm{R_c}} \mathrm{Tr}\sum_{\Lambda \Lambda'}Z_\Lambda(E, \mr) t_{\Lambda\Lambda'}(E) Z^+_{\Lambda'}(E, \mr) d\mr-\frac{4p W}{3c^2 \pi} \mathrm{R_c}^3.
\end{align}
For $r>R_c$, using another expression of the Green's function
\begin{align}
G(E, \mathbf{r},\mathbf{r})=G_0(E, \mathbf{r},\mathbf{r})-p^2\sum_{\Lambda \Lambda'}H_\Lambda(E, \mr) t_{\Lambda\Lambda'}(E) H^+_{\Lambda'}(E, \mr),
\end{align}
where $H_\Lambda(E, \mr) = J_\Lambda(E, \mr) + i N_\Lambda(E, \mr) $, the outside contribution to the DOS is
\begin{align}
n_{\mathrm{out}}(E)=\mathrm{Im}\; \frac{2 p^2 W}{\pi c^2}\int_{R_c}^{\infty}\sum_{l} h_l(pr)t_{ll}  h_l(pr) r^2dr.
\end{align}
This double Hankel function integral also occurs in the non-relativistic case \cite{YangWang} and can be done analytically using \cite{Sutthorp}
\begin{align}
\int_R^\infty h_l(pr)^2r^2dr=\frac{R^3}{2}\left( h_{l-1}(pr) h_{l+1}(pr)-h_l(pr)^2\right).
\end{align}
To compare the Krein's theorem method and the Green's function method, the DOS of copper and vanadium are calculated and shown in figure \ref{fig:GreenVSKrein}. The input potentials are obtained from non-relativistic self-consistent full-potential KKR calculations. In both calculations 256 energy points are used.   In general the relative error is within the order of $10^{-4}$, which indicates an excellent agreement between the two methods. Most of the errors are from the small number of energy points and the relative primitive way to calculate $n_{\mathrm{K}}$ from the numerical derivative of $N_\mathrm{K}$.

\section{Singel-site DOS}
\begin{figure}[h!]
\centering
   \includegraphics[width=0.67\textwidth]{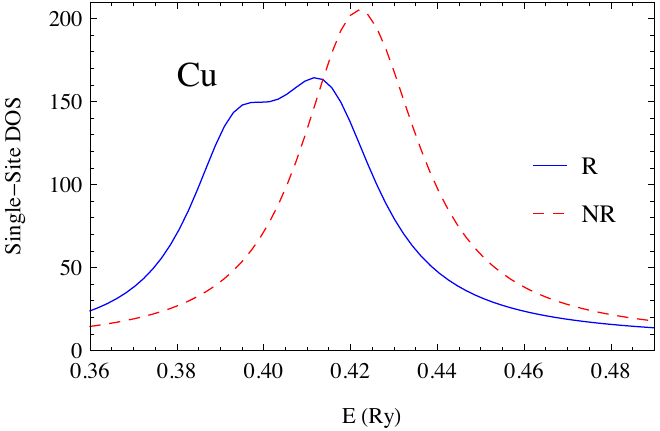}
   \includegraphics[width=0.7\textwidth]{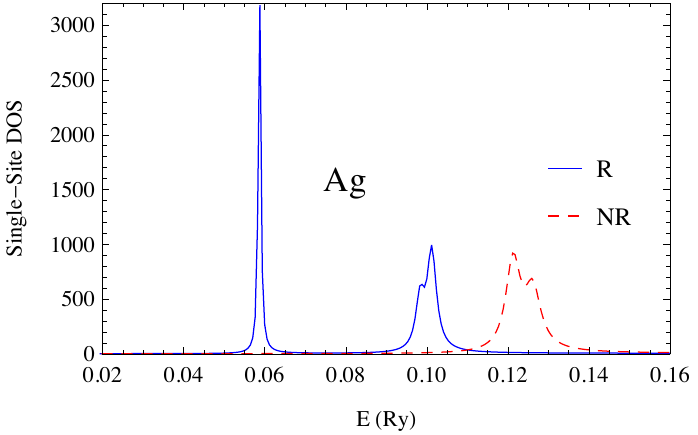}
   \includegraphics[width=0.7\textwidth]{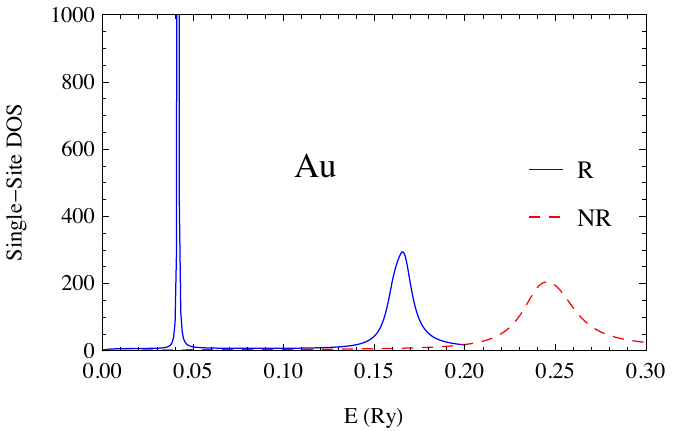}
\caption{ (Color online) Comparison of the relativistic and nonrelativistic single-site DOS of noble metals. The blue solid lines are the DOS calculated from the relativistic  full potential Green's function method. The red dashed lines are the DOS calculated from the non-relativistic full potential Green's function method, which solves the Schr\"odinger's equation. Note that the first peak of Au is not completely shown because it's too sharp. } \label{fig:DOS_Noble}
\end{figure}
The full potential effects and the relativistic effects can be directly observed in the single-site DOS. As the first example, the DOS of noble metals, i.e., copper, silver and gold are calculated using both relativistic and nonrelativistic Green's function methods. The results are shown in figure \ref{fig:DOS_Noble}.  In all calculations the angular momentum expansion cut-offs $l_{max}$ are set to be 4 and the expansion cut-offs for potentials are set to be $2 \times l_{max}$ to satisfy the angular momentum triangular relation.
Both the relativistic effects and the full potential effects  are  well demonstrated by the three peaks in the relativistic DOS plots of silver. The large energy differences between the leftmost peak and the right two peaks are due to relativistic effects, mostly spin-orbit coupling. The smaller difference between the right two peaks is due to full potential effect. The full potential effects of copper and gold, however, do not cause any visible splitting of DOS because their peaks on the right are relatively broad and merge into one single peak.
\begin{figure}[h!]
\centering
   \includegraphics[width=0.45\textwidth]{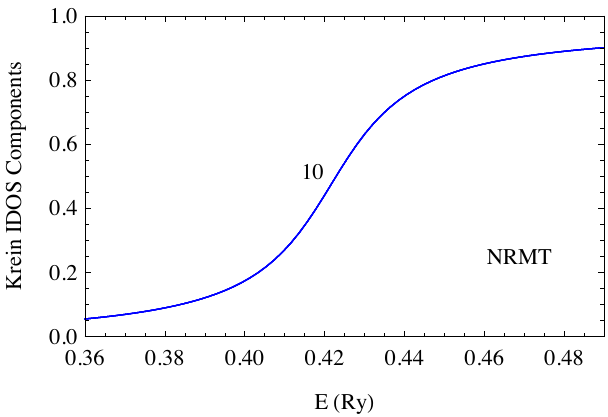}
   \includegraphics[width=0.45\textwidth]{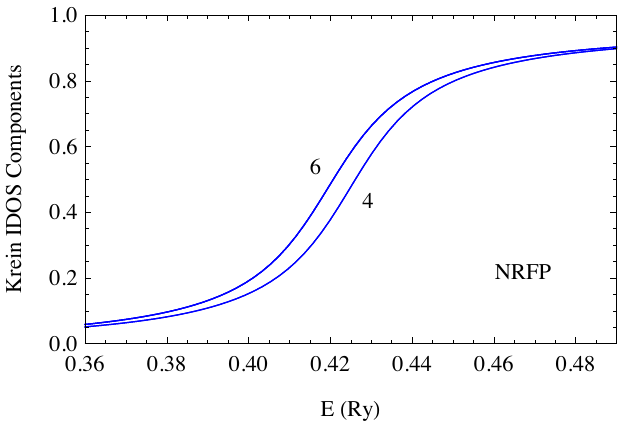}
   \includegraphics[width=0.45\textwidth]{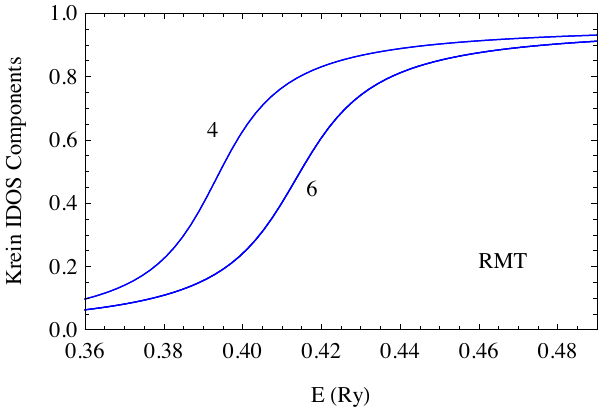}
   \includegraphics[width=0.45\textwidth]{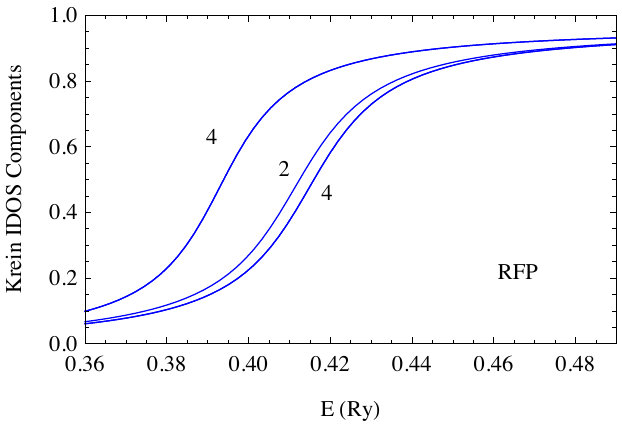}
\caption{ The partial Krein IDOS of d electrons of copper corresponding to non-relativistic muffin-tin (NRMT), non-relativistic full potential (NRFP), relativistic muffin-tin (RMT) and relativistic full potential (RFP) calculations. There are 10 d channels in total and the number of degeneracy is shown for each curve.} \label{fig:Phase}
\end{figure}
\begin{figure}[h!]
\centering
   \includegraphics[width=0.7\textwidth]{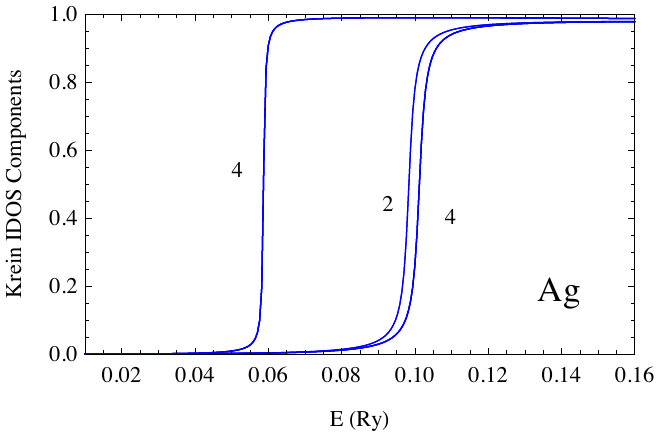}
   \includegraphics[width=0.7\textwidth]{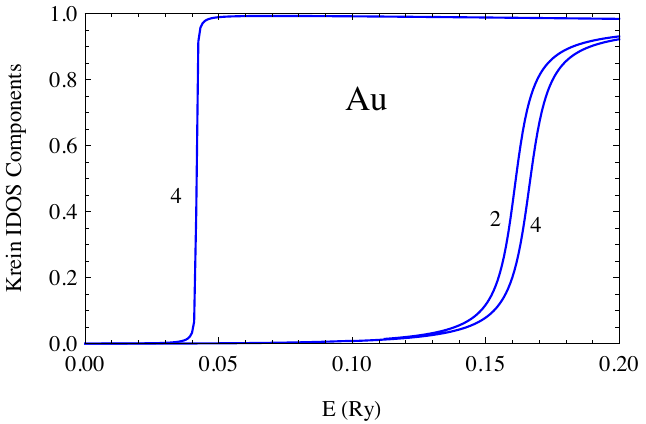}
\caption{ The partial IDOS of d electrons of silver and gold in relativistic full potential calculation. There are 10 d channels in total and the number of degeneracy is shown for each curve.  } \label{fig:Phase_AgAu}
\end{figure}
A better demonstration of the full potential effects and the relativistic effects is the splitting of the IDOS components of the d electrons. According to Krein's theorem, the IDOS components are given by the generalized phase shifts divided by 2$\pi$, where the generalized phase shifts are obtained by diagonalizing the S-matrix \cite{Gyorffy}.  As an example we calculated the Krein IDOS components of copper in different cases and the results are shown in figure \ref{fig:Phase}. For non-relativistic muffin-tin calculation, all the d electrons degenerate. For relativistic muffin-tin calculation, even though the input potential is spherically symmetric, the IDOS still splits into two parts due to spin-orbit coupling. The introduction of asymmetric potential leads to further splitting in the relativistic full potential calculation. Because of the cubic structure symmetry, the splitting is not complete and the degeneracies still exist. Similar calculations are also performed for silver and gold and the Krein IDOS for relativistic full-potential calculation are shown in figure \ref{fig:Phase_AgAu}, where the splittings due to spin-orbit coupling are more significant. The width of the spitting provides an estimate of the strength of the spin-orbit coupling or full potential effect. For example, for copper, silver and gold the splittings caused by spin-orbit coupling are 0.020 Ry, 0.042 Ry and 0.13 Ry, respectively. This agrees with our expectation that the magnitude of spin-orbit coupling tend to increase for heavier elements.
\begin{figure}[h!]
\centering
    \includegraphics[width=0.7\textwidth]{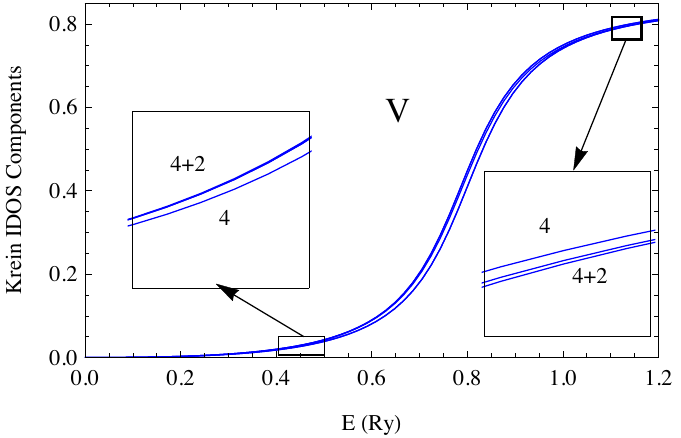}
    \includegraphics[width=0.7\textwidth]{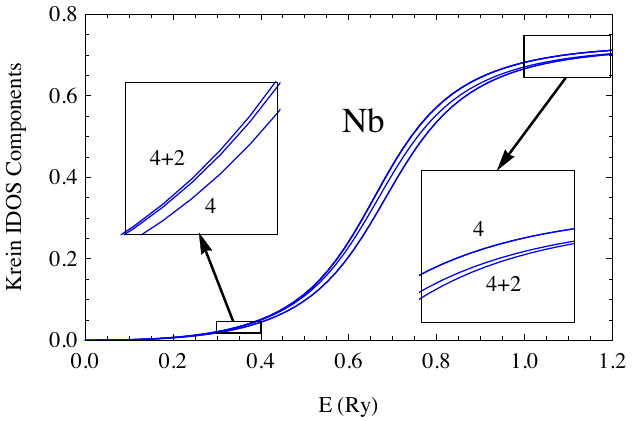}
    \includegraphics[width=0.7\textwidth]{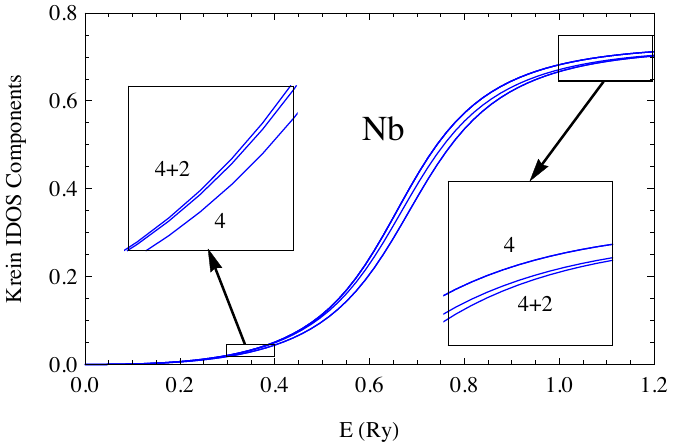}
\caption{The Krein IDOS components of the d electrons of group V elements. There are 10 d channels in total and the number of degeneracy is shown for each curve.} \label{fig:Phase_Nb}
\end{figure}

All the noble elements have face centered cubic structure, therefore the FP effects are comparatively small. To better investigate the FP effects we also calculated the group V elements, where the FP effects should be larger because of they are body centered cubic crystals and are less closely packed. The Krein IDOS components of vanadium, niobium and tantalum are shown in figure \ref{fig:Phase_Nb}. For Ta the relativistic effect is almost always dominant over the FP effect because of its large nuclear charge. For V and Nb, however, it is easy to observe the competition between FP effects and relativistic effects. At small energies, the FP effects are dominant over the relativistic effects. As a result, the 10 d components evolve into two major branches. The upper branch has approximately 6 fold degeneracy, with angular momentum index $l,m=2,0$ or $2,\pm 2$ and the lower branch has 4 fold degeneracy, with $l,m=2, \pm 1$. At high energies, however, the relativistic effects will be dominant. Although there are still two branches, due to spin orbit coupling, the good quantum number is the total angular momentum $J$. The upper branch now is 4 fold degenerate, with $J=3/2$ and the lower branch is 6  fold degenerate, with $J=5/2$. This is why in the intermediate region the two components in the middle move from the lower branch to the upper branch as the energy increase.

At the end of this section we would like to discuss some general features in the plots: The first one is that the resonance peak tends to be sharper when the center of the peak is located at lower energy. This can be best understood by imagining the energy of the d-resonance moves to negative, at that point, the resonance peak will become a delta function and this corresponds to a bound state of the system. The second one is that the resonance peak in the relativistic case tends to move toward lower energies compared to the non-relativistic one.  Taken to the extreme it means a resonance state can become a bound state due to relativistic effects. This can be seen as a result of relativistic contraction: the electrons move closer to the nucleus, which effectively expands the size of the potential, therefore more bound states can be  accommodated.

\section{Single-site Charge Density}
\begin{figure}[h!]
\centering
    \includegraphics[width=0.7\textwidth]{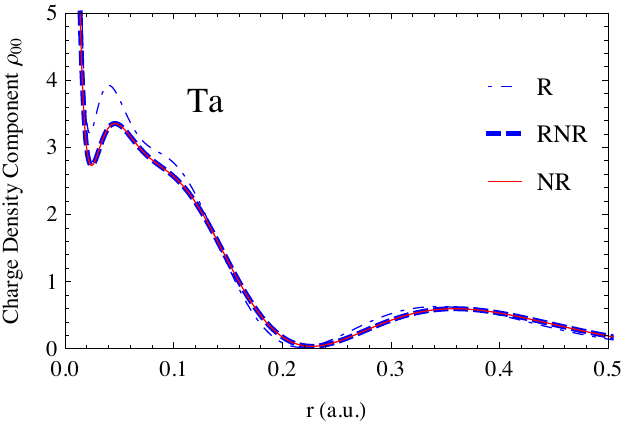}
\caption{ (Color online) The $l,m=(0,0)$ component of the valence-band charge density of tantalum calculated with relativistic (R) method, non-relativistic (NR) Green's function method that solves Schr\"odinger's equation, and relativistic method at non-relativistic limit (RNR).   } \label{fig:ChargeD00}
\end{figure}
\begin{figure}[h!]
\centering
    \includegraphics[width=0.7\textwidth]{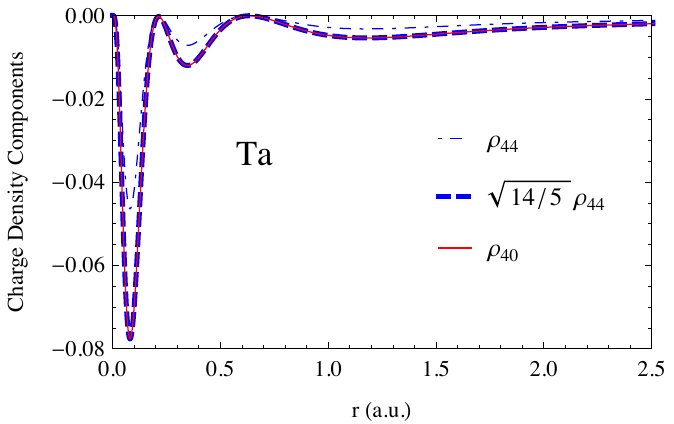}
\caption{ (Color online) The $l,m=(4,0)$ component of the valence-band charge density of copper. Comparison of $\rho_{4,0}$ and $\sqrt{14/5}\;\rho_{4,4}$ is made to test the cubic symmetry in our results.  } \label{fig:ChargeD40}
\end{figure}
In addition to the density of states, the electron charge density distribution is another quantity that is essential for a self-consistent electronic structure calculation. Again we focus our attention to single-site scattering and hence only show the single-site charge densities here. Also by "charge density" we actually mean the valence electron charge density.  Using equation (\ref{CD}), the single-site charge density can be found by integrating the Green’s function over energy.  It’s convenient to expand the charge density in terms of the spherical harmonics
\begin{align}
\rho(\mr)=\sum_{l=0}^{2\times l_{\mathrm{max}}}\sum_{m=-l}^l \rho_{l,m}(r) Y_{l,m}(\theta,\phi).
\end{align}
Because of the cubic symmetry in the elements we calculated, for $2 \times l_{\mathrm{max}}=8$ the only non-vanishing components are $\rho_{0,0}, \rho_{4,0}, \rho_{4,\pm4},\rho_{6,0},\rho_{6,\pm4},\rho_{8,0},\rho_{8,\pm4}$ and $\rho_{8,\pm8}$. Moreover, not all of the charge density components are independent \cite{Rusanu}. For example, we should have $\rho_{4,4}=\rho_{4,-4}=\sqrt{5/14}\; \rho_{4,0}$ for all cubic crystal systems. Checking the symmetry structure of  the charge density also serves as a verification of our results.

\begin{figure}[h!]
\centering
    \includegraphics[width=0.7\textwidth]{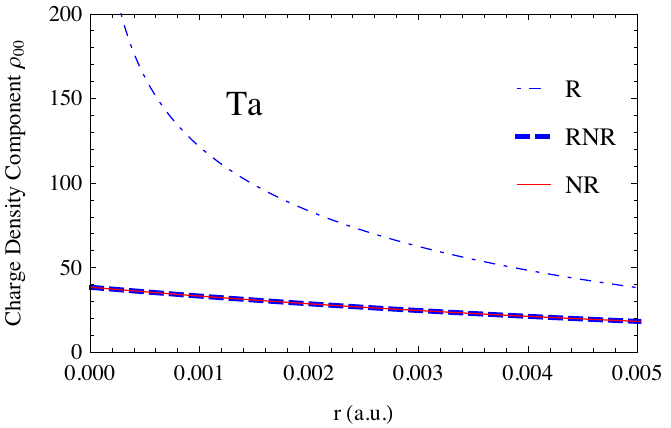}
\caption{ (Color online) Comparison of the relativistic (R) and nonrelativistic (NR) charge density around the origin. The non-relativistic limit of our relativistic method (RNR) is also shown.  } \label{fig:Cusp}
\end{figure}

As an example, we calculated the charge density components of tantalum, and the proportional relationship of $\rho_{4,0}$ and $\rho_{4,4}$ is shown in figure \ref{fig:ChargeD40}, with relative error of the ratio found to be within $10^{-7}$. We  also studied the relativistic effects by comparing the relativistic and nonrelativistic charge density components $\rho_{0,0}$ as shown in figure \ref{fig:ChargeD00}. The non-relativistic limit of our method is also taken by  setting $1/c=0$ and compared with the non-relativistic Green's function method to verify our results. From figure \ref{fig:ChargeD00} we see the relativistic charge density contracts towards the origin compared to the non-relativistic one, which is a typical relativistic phenomenon \cite{Burke}. Taking a closer look, the relativistic contraction is best demonstrated by the asymptotic behavior of the charge density near the origin, as shown in figure \ref{fig:Cusp}.  The non-relativistic  $\rho_{0,0}\approx A(1-\alpha r)$ \cite{Rusanu} near the origin, with $\alpha=2Z$; The relativistic $\rho_{0,0}$, however, demonstrates a weak singular behavior.  This is no surprise since it's well known that when the nucleus is considered as a point charge, the relativistic $S_{1/2}$ and $P_{1/2}$ partial waves will behave as $r^{\sqrt{\kappa^2-\zeta^2}-1}$ around the origin\cite{XDWang}, where $\zeta=2Z/c$. Note that this divergence is not pathological because when integrated around origin,  $\rho_{0,0}$ will be multiplied by $r^2$, which leads to a finite value. As a final check of our results, we compared the magnitude of the charge density components and find the total charge density is indeed positive everywhere.

\section{Conclusions}
We have demonstrated the construction of the RFP single-site Green's function using the scattering matrices and solutions obtained from directly solving the Dirac equations. Compared to other approaches, our method employs the sine and cosine matrices formalism. As a result, no matching on the boundary is needed for solving the Dirac equation. 

This RFP method is successfully implemented to calculate the Green's functions.  To test the code, the single-site DOS of noble metals and group V elements are calculated using both the Krein's theorem method and the Green's function method and an excellent agreement between the two is found. By studying the single-site DOS and the scattering phase shifts, we thoroughly investigated the relativistic effects and the full-potential effects in single-site scattering. Finally, using tantalum as an example,  we investigated the relativistic contraction of the charge density and the asymptotic behavior of charge density near the origin.

Currently the input potential of our code is provided by another self-consistent non-relativistic KKR calculation. To develop a MST based relativistic ab initio
electronic structure calculation program, we still need to calculate the new potential from the electron charge density distribution.  This part is essentially the same as the non-relativistic calculation, therefore this work provides a good foundation to develop a MST based relativistic ab initio electronic structure calculation program.
\section{Acknowledgements}
This work has been sponsored by the U.S. Department of Energy, Office
of Science, Basic Energy Sciences, Material Sciences and Engineering
Division (M.E. and the later work of G.M.S) and by the Center for Defect
Physics, an Energy Frontier Research Center of the Office of Basic Energy
Sciences of the U. S. Department of Energy.

\bibliographystyle{unsrt}
\bibliography{ref01}

\appendix
\section{Derivation of the solutions $\boldsymbol{\psi}(\mr) $}
In this appendix we show how the expression in equation (\ref{eq03}) is obtained. The free-space Green’s function $G_0(E, \mr,\mr')$ of the Dirac equation can be expressed in terms of the free-space solutions,
\begin{align}
G_0(E, \mr,\mr')=&-ip\sum_\Lambda J_\Lambda (E, \mr) H_\Lambda^+ (E, \mr') \Theta(\mr'-\mr) \nonumber \\ 
 &-ip\sum_\Lambda H_\Lambda (E, \mr) J_\Lambda^+ (E, \mr') \Theta(\mr-\mr').\label{freeG}
\end{align}
The solutions of the Dirac equation can be written in terms of the free-space Green’s function $G_0(E, \mr,\mr')$ and the free-space solutions $\phi(\mr)$:
\begin{align} \label{eq01}
\psi(E, \mr)=\phi(E, \mr)+\int_{\Omega} d^3\mr' G_0(E, \mr,\mr') V(\mr')\psi(E, \mr').
\end{align}
 We have the freedom of choosing $\phi(\mr)$. Here our choice is:
\begin{align}
\phi(E, \mr)=\sum_{\Lambda'} {J}_{\Lambda'}(E, \mr)\{ i S_{\Lambda' \Lambda}(E)-C_{\Lambda' \Lambda}(E) \}, \label{eqphi}
\end{align}
where
\begin{align}
C_{\Lambda' \Lambda}(E)  =& p \int_\Omega d^3\mr' N_{\Lambda'}^{+}(E, \mr') V(\mr')\psi_{\Lambda}(E, \mr')-\delta_{\Lambda' \Lambda} \label{CM}\\
S_{\Lambda' \Lambda}(E) =& p \int_\Omega d^3\mr' J_{\Lambda'}^{+}(E, \mr') V(\mr')\psi_{\Lambda}(E, \mr'). \label{SM}
\end{align}
equation (\ref{eq01}) becomes
\begin{align} \label{eq02}
\psi_\Lambda(E, \mr)=&\sum_{\Lambda'} \{ i S_{\Lambda' \Lambda}(E) -C_{\Lambda \Lambda'}(E) \} J_{\Lambda'}(E, \mr)+\nonumber \\
&\int_{\Omega} d^3\mr' G_0(E,\mr,\mr') V(\mr')\psi_\Lambda(E, \mr').
\end{align}
Next we plug the free-space Green’s function expression equation (\ref{freeG}) back into equation (\ref{eq02}) and split the integral region into two pieces according to $r'>r$ or $r'<r$. After some manipulation,  we obtain equation (\ref{eq03}). i.e.
\begin{align}
\psi_\Lambda(E, \mr)=\sum_{\Lambda'} \{ S_{\Lambda' \Lambda}(E, r) N_{\Lambda'}(E, \mr) -C_{\Lambda' \Lambda }(E, r) J_{\Lambda'}(E, \mr) \},
\end{align}

\section{Derivation of the Green's function}
In this appendix we show how to derive the expression of Green's function in equation (\ref{Green}). The general expression of Green’s function is
\begin{align}
G(E, \mr,\mr')=&-ip\sum_\Lambda P_\Lambda(E, \mr ) Q^+_\Lambda(E, \mr')\Theta(\mr'-\mr) \nonumber \\
&-ip \sum_\Lambda Q_\Lambda(E, \mr ) P^+_\Lambda(E, \mr')\Theta(\mr-\mr').
\label{eq70}
\end{align}
We have obtained regular solutions $\psi_{\Lambda}(E, \mr)$. To find the appropriate irregular solutions to construct the Green's function, we start with the definition of the Green's function 
\begin{align}
\left[-c\boldsymbol{\alpha} i \nabla+\beta mc^2+V(\mr)-W\right]G(E, \mr,\mr')=-\delta(\mr-\mr').
\end{align}
Simplify the above expression, we find
\begin{align}
(-i)\oint_{r=R2}d\mathbf{s}\cdot \mathbf{W}[P^+_\Lambda(E, \mr), Q_{\Lambda'}(E, \mr)]=\delta_{\Lambda \Lambda'}, \label{WPQ}
\end{align}
where $\mathbf{W}[\cdots]$ is the relativistic Wronskian
\begin{align}
\mathbf{W}[\psi_1^+(E, \mr),\psi_2(E, \mr)]=icp\psi_1^{+}(E, \mr) \boldsymbol{\alpha} \psi_2(E, \mr).
\end{align}
The following Wronskian relations satisfied by the free-space solutions can be verified by pluging in equation (\ref{J}), (\ref{N}),(\ref{Jplus}) and (\ref{Nplus}),
\begin{align}
\oint_{r'=r}d\mathbf{s}'  \mathbf{W}[J^+_\Lambda(E, \mr'),  N_{\Lambda'}(E, \mr')] &=-\oint_{r'=r}d\mathbf{s}'  \mathbf{W}[N^+_\Lambda(E, \mr'),  J_{\Lambda'}(E, \mr')] =\delta_{\Lambda \Lambda'}, \\
\oint_{r'=r}d\mathbf{s}'  \mathbf{W}[J^+_\Lambda(E, \mr'),  J_{\Lambda'}(E, \mr') ]&=0,
\\
\oint_{r'=r}d\mathbf{s}'  \mathbf{W}[N^+_\Lambda(E, \mr'), N_{\Lambda'}(E, \mr')] &=0.
\end{align}
For a given regular solution, equation (\ref{WPQ}) is the relation the irregular solution need to satisfy to construct the Green's function in equation (\ref{eq70}). To obtain an explicit expression of $Q_\Lambda(E,\mr ) $, we make use of the integral equation satisfied by the Green's functions
\begin{align}
G(E, \mr,\mr')=G_0(E, \mr,\mr')+\int_\Omega d^3\mr'' G_0 (E, \mr,\mr')V(\mr'') G(E, \mr'',\mr').
\end{align}
Plug in equation (\ref{eq70}) and equation (\ref{freeG}) and use the Dirac equations,  it can be shown that the explicit expression of the irregular solution is given by
\begin{align}
Q^+_\Lambda(E, \mr)=\sum_{\Lambda'}\left[ A^{-1}_{\Lambda\Lambda'}(E)\right] \mathcal{H}^+_{\Lambda'}(E, \mr),
\end{align}
where $\mathcal{H}^+_{\Lambda}(\mr') $ have similar definition as  $\mathcal{J}^+_{\Lambda}(E, \mr') $, with the boundary conditions $\mathcal{H}^+_{\Lambda'}(E, \mr) =H^+_{\Lambda'}(E, \mr) $  outside the Wigner-Seitz cell,  and
\begin{align}
A_{\Lambda'\Lambda}(E)=-i\oint_{B_\Omega} d\mathbf{s}''\cdot \mathbf{W}[\mathcal{H}^+_{\Lambda}(E, \mr''),P_\Lambda(E, \mr'')].
\end{align}
If we plug in equation (\ref{eq03}) and the relativistic Wronskian relations satisfied by the free space solutions,
we can find the corresponding irregular solution to construct the Green’s function to be
\begin{align}
Q^+_\Lambda(E, \mr)=\sum_{\Lambda'}[iS_{\Lambda'\Lambda}(E)-C_{\Lambda'\Lambda}(E)]^{-1} \mathcal{H}^+_{\Lambda'}(E, \mr).
\end{align}
Therefore, for $r'>r$ the Green's function is given by the expression
\begin{align}
G(E, \mr,\mr')=-ip\sum_{\Lambda \Lambda'}\psi_\Lambda(E, \mr)[iS_{\Lambda'\Lambda}(E)-C_{\Lambda'\Lambda}(E)]^{-1} \mathcal{H}^+_{\Lambda'}(E, \mr').
\end{align}
To obtain a more familiar expression, we use the relation satisfied by the $T$-operator
\begin{align}
\langle J^+_\Lambda(E, \mr)|\hat{V}|\psi_{\Lambda'}(E, \mr)\rangle=\langle J^+_\Lambda(E, \mr)|\hat{T}|\phi_{\Lambda'}(E, \mr)\rangle.
\end{align}
Plug in equation (\ref{eqphi})  and (\ref{eq03}), it's easy to show that the $t$ matrix is given by
\begin{align}
t_{\Lambda \Lambda'}(E)^{-1}=p\left(i\delta_{\Lambda \Lambda'}-\sum_{\Lambda''}C_{\Lambda \Lambda''}(E) S^{-1}_{\Lambda'' \Lambda'}(E)\right).
\label{TInv}
\end{align}
Now use the definition of $Z_\Lambda(E, \mr)$ in equation (\ref{Z}), we obtain the useful expression
\begin{align}
G(E, \mr,\mr')&=-ip\sum_{\Lambda \Lambda'}Z_\Lambda(E, \mr) t_{\Lambda\Lambda'}(E) \mathcal{H}^+_{\Lambda'}(E, \mr') \\
&=\sum_{\Lambda \Lambda'}Z_\Lambda(E, \mr) t_{\Lambda\Lambda'}(E) Z^+_{\Lambda'}(E, \mr')-\sum_{\Lambda} Z_\Lambda(E, \mr) \mathcal{J}^+_{\Lambda}(E, \mr')    .
\end{align}
In the second line, we used the relation
\begin{align}
Z^+_{\Lambda'}(E, \mr)=\sum_{\Lambda''} \left[t_{\Lambda'\Lambda''}(E)^{-1} \right]^+ \mathcal{J}^+_{\Lambda''} (E, \mr) -ip\mathcal{H}^+_{\Lambda'}(E,\mr),
\end{align}
and the fact that
\begin{align}
\left[t_{\Lambda'\Lambda''}(E) \right]^+=\left[t_{\Lambda'\Lambda''}(E)\right]^T,
\end{align}
which can be proved using relativistic Wronskians by analogy with the non-relativistic case \cite{Rusanu}.

Because both the potential and the energies are real, from equation (\ref{dSdr}-\ref{dd}), we see that for our calculation $S_{\Lambda' \Lambda}(E, r)$ and $C_{\Lambda' \Lambda}(E, r)$ are also real, from the Dirac equation satisfied by the left-hand solutions, we have
\begin{align}
\psi^+_\Lambda(E, \mr)=\sum_{\Lambda'} \{  N^+_{\Lambda'}(E, \mr)  \left[ S_{\Lambda' \Lambda}(E, r)\right]^T- J^+_{\Lambda'}(E, \mr) \left[ C_{\Lambda' \Lambda }(E, r)\right]^T \},
\end{align}
and
\begin{align}
Z^+_\Lambda(E, \mr)=p\sum_{\Lambda'} \psi^+_{\Lambda'} (E,\mr)\left[S^{-1}_{\Lambda' \Lambda}(E) \right]^{T}. \label{Zplus}
\end{align}
The ``$T$'' superscript indicates taking the transpose. Note that when the energy is complex, the left-hand solutions and the right-hand solutions do not have such a simple relation, i.e.,  we cannot obtain $\left[ S_{\Lambda' \Lambda}(E, r)\right]^+$ by taking the transpose of $S_{\Lambda' \Lambda}(E, r)$, but we can still obtain one from the other by using the relativistic generalization of the ``dot'' notation in \cite{Rusanu}. In the most general case that the potential is non-Hermitian and the energy is complex, we may need to solve the left-hand and right-hand solutions individually.  Details of the distinction between right-hand and left-hand solutions at a general potential have been discussed in a recent paper \cite{RLS}.
\end{document}